\documentclass{article}
\usepackage{spconf}
\usepackage{microtype}
\usepackage{amsmath,amssymb,amsfonts,mathtools}
\usepackage{booktabs}
\usepackage{multirow}
\usepackage{tabularx}
\usepackage{array}
\usepackage{float}
\usepackage{adjustbox}
\usepackage{xcolor}
\usepackage{url}
\usepackage[numbers,sort&compress]{natbib}
\usepackage[hidelinks]{hyperref}
\newcolumntype{C}{>{\centering\arraybackslash}X}
\graphicspath{{./}{figures/}}
\title{SHB-AE: Spherical harmonic beamforming based Ambisonics encoding and upscaling method for smartphone microphone array}
\name{Yuhuan You\textsuperscript{1}, Yufan Qian\textsuperscript{1}, Tianshu Qu\textsuperscript{1}, Bin Wang\textsuperscript{2}, Xueyang Lv\textsuperscript{3}}
\address{$^{1}$Peking University\\
$^{2}$Beijing Xiaomi Mobile Software Co., Ltd\\
$^{3}$Xiaomi Communications Co., Ltd}

\begin{document}

\maketitle

\begin{center}
\scriptsize\itshape
Accepted for presentation at\\
AES Europe 2025 Convention\\
(AES 158th Convention)\\
Warsaw, Poland, May 22-24, 2025.\\
This is the authors' version prepared\\
without AES graphical style or template.
\end{center}

\begin{abstract}
With the rapid development of virtual reality (VR) and augmented reality (AR), spatial audio recording and reproduction have gained increasing research interest. Higher Order Ambisonics (HOA) stands out for its adaptability to various playback devices and its ability to integrate head orientation. However, current HOA recordings often rely on bulky spherical microphone arrays (SMA), and portable devices like smartphones are limited by array configuration and number of microphones. We propose SHB-AE, a spherical harmonic beamforming based method for Ambisonics encoding using a smartphone microphone array (SPMA). By designing beamformers for each order of spherical harmonic functions based on the array manifold, the method enables Ambisonics encoding and up-scaling. Validation on a real SPMA and its simulated free-field counterpart in noisy and reverberant conditions showed that the method successfully encodes and up-scales Ambisonics up to the fourth order with just four irregularly arranged microphones.
\end{abstract}

\section{Introduction}
With the rapid development of augmented reality (AR) and virtual reality (VR) technologies, the demand for immersive spatial audio experiences has been continuously rising, leading to growing attention on spatial audio capture and reproduction technologies. Among these, encoding microphone-captured sound pressure signals into Higher-Order Ambisonics (HOA) format offers multiple advantages\cite{zotter2019ambisonics}. HOA signals are device-independent and can be used in various playback environments, such as headphones and loudspeaker arrays. Additionally, HOA signals can be conveniently implemented with head rotation and the head-related transfer function (HRTF) through the Wigner-D function and spherical harmonic coefficients \cite{Magariyachi_Mitsufuji_2020}.

However, existing Ambisonics encoding algorithms typically rely on bulky spherical microphone arrays to capture sound pressure signals \cite{zotter2019ambisonics}, making them difficult to adapt to portable electronic devices not specifically designed for spatial sound capture, such as smartphones, glasses, and other wearable devices \cite{Talagala_Zhang_Abhayapala_2013}. These devices often have irregular microphone arrangements, limited microphone counts, and complex sound scattering, which limits accurate Ambisonics encoding \cite{rafaely2015fundamentals}. 

To address these challenges, several approaches have been proposed for Ambisonics encoding with irregular arrays \cite{heikkinen2024neural,qiao2024neural,McCormack_parametric,Abhayapala_independetn,gayer2024ambisonics}. These methods can be broadly classified into data-driven and model-driven approaches. Heikkinen et al. \cite{heikkinen2024neural} introduced a data-driven method for Neural Ambisonics Encoding, where a U-net structured neural network is trained to directly encode Ambisonics from sound pressure signals captured by irregular microphone arrays in an end-to-end manner. Building on this, Qiao et al. \cite{qiao2024neural} regularized inter-channel correlations of the Ambisonics signals and utilized a circular microphone array for Ambisonics encoding of multi-speaker speech signals. However, like many data-driven methods, these approaches face challenges related to generalization errors in the absence of actual microphone-captured signals, which are difficult to obtain.

One type of model-driven approach is based on analyzing the captured sound field. For example, McCormack et al. \cite{McCormack_parametric} introduced a primary-ambience directional model of the sound field, using spatial filtering techniques to divide the captured sound field into its individual source and directional ambient components, which are then encoded into the Ambisonics format at an arbitrary order. However, these methods rely not only on prior assumptions about the sound field but also on the impact of direction-of-arrival (DOA) estimation and source separation results. Accurate DOA estimation and source separation with SPMA is a challenging task, and most studies assume high-precision HOA recordings as input for these tasks \cite{wdh_trans,wdh_nature}. 

Another type of model-driven approach is signal-independent encoding, which directly formulates a system of equations linking HOA coefficients and captured sound pressure signals \cite{Abhayapala_independetn}, solving for the optimal encoding matrix in the least squares sense. Bastine et al. \cite{bastine2022ambisonics} directly formulated a linear system of equations between HOA coefficients and the array manifold of head-wearable devices to achieve encoding for microphone arrays on such devices. Geyer et al. \cite{gayer2024ambisonics} further introduced Tikhonov regularization \cite{Golub_Hansen_O’Leary_1999} to improve the performance of encoded HOA signals in binaural rendering tasks. However, these methods are still constrained by the number of microphones $Q$ in the array, with the limitation $Q\geq (N+1)^2$ \cite{rafaely2015fundamentals}. Gao et al. \cite{gao2018high}, correspondingly, designed beamformers to achieve Ambisonics encoding, utilizing 2.5D spherical harmonic functions as target beam patterns.

We propose SHB-AE, a beamforming-based method for Ambisonics encoding and upscaling. Inspired by prior beamforming work \cite{gao2018high}, we design beamformers corresponding to spherical harmonic functions as target beam patterns, enabling direct output of HOA coefficients for each order without being limited by microphone array configuration or microphone numbers. The limitation of microphone numbers is converted into the limitation of the number of measured steering vectors, which allows the up-scaling of Ambisonics encoding. Experimental results demonstrate superior performance over conventional least-square HOA encoding across various reverberant environments and noise conditions.

The remainder of the paper is organized as follows. Section 2 introduces the fundamental knowledge of Ambisonics encoding and the proposed method. Section 3 presents experimental validation of the method’s effectiveness across simulated and measured array manifolds of an SPMA under multiple conditions. Section 4 concludes the paper.

\section{Methods}
\subsection{Ambisonics Encoding}
\label{section:Ambisonics}

Ambisonics is a spatial audio technique that allows for the reproduction of three-dimensional sound fields using spherical harmonics. This method enables immersive sound experiences by representing sound from various directions, which can be accurately reproduced across different listening environments.

Consider a plane wave $e^{i \mathbf{k} \cdot \mathbf{r}_q}$ in spherical coordinates where $\mathbf{k} = (k, \theta_k, \phi_k)$ stands for a plane wave arriving from $(\theta_k, \phi_k)$, the sound pressure $p(k,\mathbf{r}_q)$ at the $q$-th microphone $\mathbf{r}_q = (r_q, \theta_q, \phi_q)$ is \cite{rafaely2015fundamentals}
\begin{equation}
\label{eq:basic}
\begin{aligned}
    e^{i \mathbf{k} \cdot \mathbf{r}_q} 
    &= p(k, r_q, \theta_q, \phi_q)\\
    &= 4\pi\sum_{n=0}^N \sum_{m=-n}^n b_n(kr_q)B_n^m(\theta_k, \phi_k) Y_n^m(\theta_q, \phi_q)
\end{aligned}
\end{equation}
where $b_n(kr)=i^nj_n(kr)$, \( j_n(kr) \) denotes the spherical Bessel function. When the truncation order $N \to \infty$, the above equation holds. The spherical harmonics \( Y_n^m(\theta, \phi) \) are defined as
\begin{equation}
    Y_n^m(\theta, \phi) \equiv \sqrt{\frac{2 n+1}{4 \pi} \frac{(n-m)!}{(n+m)!}} P_n^m(\cos \theta) \mathrm{e}^{\mathrm{i} m \phi}
\end{equation}
with \( P_n^m \) being the associated Legendre functions. Within this framework, the theoretical Ambisonics coefficient \( B_n^m(\theta_k, \phi_k) \) is given by
\begin{equation}\label{eq:theor_hoa}
    B_n^m(\theta_k, \phi_k) = [Y_n^m(\theta_k, \phi_k)]^*
\end{equation}
In practice, Ambisonics signals are obtained using a microphone array and projected onto spherical harmonic bases as
\begin{equation}\label{eq:encode}
    B_n^m = \frac{1}{4\pi} \sum_{q=1}^Q w_q p(\theta_q, \phi_q) [Y_n^{m}(\theta_q, \phi_q)]^*
\end{equation}
where $Q$ is the number of microphones, $q=1,\cdots,Q$ and $w_q$ denote the microphone index and its sampling weight respectively.

In our simulation and practical scenarios, $Q=4$, while $p, b_n(kr_q), Y_n^m(\theta_k, \phi_k)$ can be calculated from available data. Therefore, for each sound source direction $(\theta_k, \phi_k)$, we can use Equation \ref{eq:basic} to express the sound pressure at all four microphones, thus establishing a linear equations system for the coefficients $B_n^m$ of different orders. When the truncation order $N=1$ (First Order Ambisonics), the system is well-determined with an exact solution, though severely affected by spatial aliasing. For higher truncation orders, the system becomes underdetermined, thus we employ pseudo-inverse to obtain the minimum norm solution
\begin{equation}
\hat{B}_n^m = (\mathbf{A}^\dagger \mathbf{p})_{(n,m)},
\end{equation}
where $\mathbf{p}$ is the sound pressure vector. $\mathbf{A}^\dagger$ denotes the Moore-Penrose pseudo-inverse of $\mathbf{A}$, serving as the Ambisonics encoding matrix, while matrix $\mathbf{A}$ has elements 
\begin{equation}
    A_{q,(n,m)} = 4\pi b_n(kr_q)Y_n^m(\theta_q, \phi_q)
\label{eq:A_Encode_thr}
\end{equation}

\subsection{Beamforming Based Ambisonics Encoding}\label{subsec:proposed}

Consider an array of $Q$ omni-directional microphones at $(r_q, \theta_q, \phi_q)$ and a set of K plane waves arriving from $(\theta_k, \phi_k)$, the array manifold matrix is denoted as $\mathbf{D}(\omega)$ with dimensions $Q\times K$, where each element $[\mathbf{D}(\omega)]_{q,k}$ represents the frequency response of the plane wave source $k$ to microphone $q$. The sound pressure measured by the array can be expressed as
\begin{equation}\label{eq:signalModel}
    \mathbf{p}(\omega) = \mathbf{D}(\omega)\mathbf{s}(\omega) + \mathbf{n}(\omega),
\end{equation}
where $\mathbf{p}(\omega) = [p_1(\omega),\cdots,p_Q(\omega)]^T$ is the sound pressure vector, $\mathbf{s}(\omega) = [s_1(\omega),\cdots,s_K(\omega)]^T$ is the source signal vector and $\mathbf{n}(\omega) = [n_1(\omega),\cdots,n_Q(\omega)]^T$ is the noise vector. Through applying beamforming to the signals, the beamformer outputs
\begin{equation}\label{eq:beamforming}
Z(\omega) = \mathbf{h}(\omega)\mathbf{p}(\omega),
\end{equation}
where $\mathbf{h}(\omega) = [h_1(\omega),\cdots,h_Q(\omega)]$ is the beamforming weight vector.
Comparing equation (\ref{eq:beamforming}) and equation (\ref{eq:encode}) leads to the finding that Ambisonics encoding can be done through designing a beamformer with the target beam pattern being the spherical harmonics. Neglecting the noise vector, vectoring the spherical harmonics $\mathbf{y}_n^m = [Y_n^m(\theta_1,\phi_1),\cdots,Y_n^m(\theta_K,\phi_K)]$, substituting equation (\ref{eq:encode}) into equation (\ref{eq:beamforming}) and omitting source signal $\mathbf{s}(\omega)$ reveals that 
\begin{equation}\label{eq:bfEncode}
    \mathbf{h}(\omega)\mathbf{D}(\omega) = \mathbf{y}_n^m.
\end{equation}
Furthermore, although the $Q$ microphones are not necessarily distributed evenly on a sphere, the steering vectors (each column of $\mathbf{V}$) for $K$ arriving plane waves can be measured following a specific spherical distribution, analogous to the measurement process of head-related transfer functions. This spatial distribution enables us to apply discrete spherical harmonic transform (DSHT) \cite{rafaely2015fundamentals} to simplify the right side of equation (\ref{eq:bfEncode}) into a one-hot vector as
\begin{equation}\label{eq:SHDbfEncode}
    \mathbf{h}(\omega)\mathbf{D}(\omega)\mathbf{S}^T = [0,\cdots,\underbrace{1}_{l=n^2+n+m+1} ,\cdots,0]^T,
\end{equation}
where $\mathbf{S}$ is the DSHT matrix in \cite{rafaely2015fundamentals}, consisting of spherical harmonic function values and sampling weights based on sampling scheme, and $l=n^2+n+m+1$ stands for the one dimensional index of spherical harmonics. Since DSHT projects the signal on the basis of spherical harmonics, the orthogonality of spherical harmonics ensures the validity of the formula  $\mathbf{y_n^m S}=[0,\cdots,1,\cdots,0]^T$. 
Through these derivations, we transform the Ambisonics encoding problem into a simple beamformer design problem. Moreover, this approach is not restricted by specific spherical harmonic orders. For irregular arrays such as SPMA, the complexity of the array manifold can potentially be exploited to achieve higher-order encoding with a limited number of beamforming weights.

As mentioned in Equation \ref{eq:theor_hoa}, which is precisely our beamforming target, we can directly obtain the HOA coefficients for plane waves from any incident direction through the beamforming filter:
\begin{equation}
\hat{\mathbf{B}}_{n}^m=\mathbf{p}\mathbf{h}^{*}
\label{HOA calc}
\end{equation}
After determining the $(n_{s},m_{s})$-order beamformer, we project the input signal through the obtained beamformer to acquire the Ambisonics component of that order. This process needs to be repeated for each order of Ambisonics, and by combining the results from all orders, we complete the Ambisonics encoding of the entire sound field.

Nevertheless, by incorporating spatial aliasing considerations, we draw inspiration from time-domain alignment techniques in binaural rendering \cite{evans1998analyzing}, extending this concept to high-frequency array manifolds. Our approach performs time-domain alignment through replacing the original array manifold with its absolute value above the frequency threshold $\omega_{th}$, which essentially preserves magnitude responses while suppressing erroneous inter-microphone phase differences. This strategy achieves dual benefits: 1) Focusing beamformer optimization on compensating amplitude distortions given the array's inherent performance limits; 2) Reducing the spherical harmonic order required for manifold representation, thereby mitigating spatial aliasing errors.

We hypothesize that the observed anti-aliasing improvement stems from exploiting the array's local density advantage - though traditional spherical harmonic decomposition fails to leverage this benefit due to numerical ill-conditioning in global surface fitting, the localized phase alignment in our method inherently utilizes the higher spatial sampling density within subarray regions. However, such performance gains are likely direction-dependent, as the effective local density varies with the direction of arrival (DOA). This directionality aspect warrants further investigation to quantify how specific DOA patterns interact with the array's geometric configuration.

For implementation, after determining $\omega_{th}$ based on array geometry, we replace high-frequency manifold components (above $\omega_{th}$) with their absolute values. While this operation reduces simulated plane-wave manifolds to constant magnitudes (by eliminating pure phase variations), it effectively preserves real-world measured manifold characteristics where scattering-induced amplitude variations remain meaningful.

\section{Experiments}

\subsection{Configuration}

\begin{figure}[t]
\centering
\includegraphics[width=0.86\linewidth]{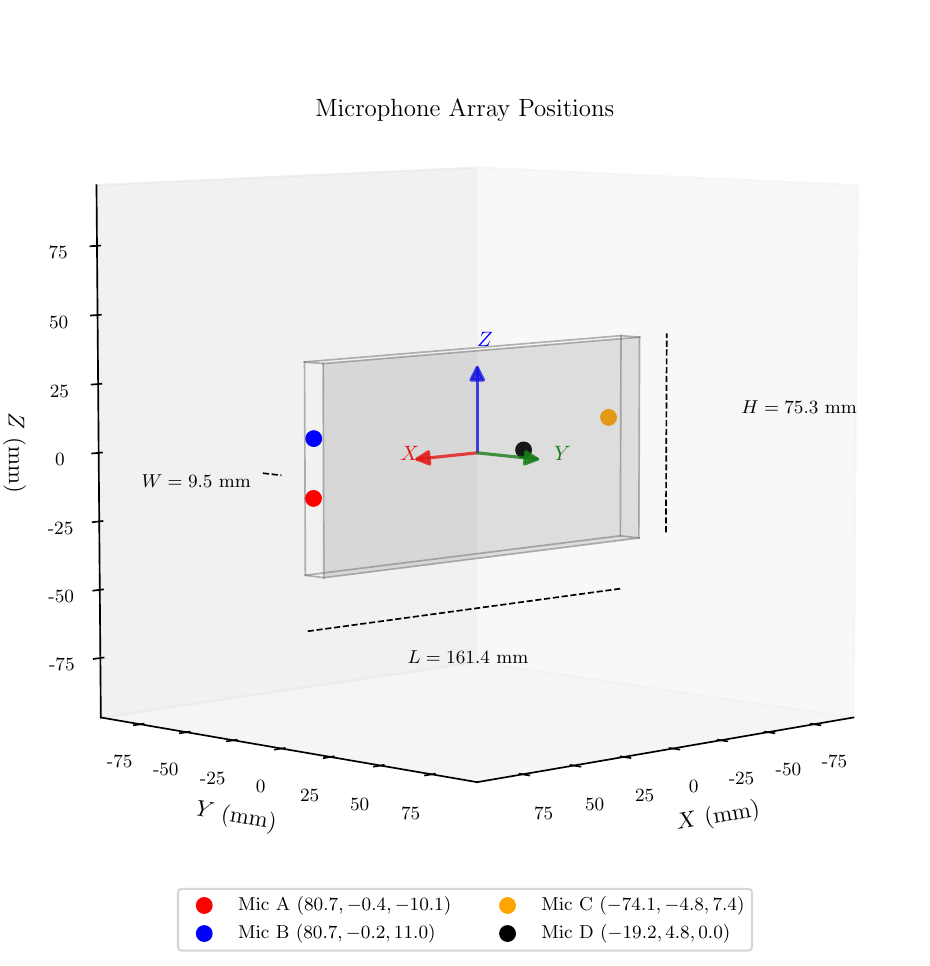}
\caption{Microphone Array Configurations of the SPMA utilized.}
\label{fg:arraypos}

\end{figure}

In our application scenario, four microphones are asymmetrically positioned on a mobile phone: two at the bottom (one primary and one secondary microphone), one at the top, and one on the rear surface. Their specific coordinates and relative positions are illustrated in Fig. \ref{fg:arraypos}. 

For filter design, as described in \citep{rafaely2015fundamentals}, we selected 50 source directions following a Gaussian distribution to recover HOA coefficients up to the 4th order ($K=2(N+1)^2$): elevation angles $\theta = \{25.02^{\circ}, 57.42^{\circ}, 90.00^{\circ}, 122.58^{\circ}, 154.98^{\circ}\}$, with azimuth angles $\phi$ sampled at $36^{\circ}$ intervals.

\subsection{Evaluation Metrics}

\subsubsection{Spatial Correlation \& Intensity Difference}

Following \cite{bertet20063d}, we evaluate the spatial correlation $R$:
\begin{equation}
R=\frac{\langle\mathbf{B}, \hat{\mathbf{B}}\rangle}{\|\mathbf{B}\| \cdot\|\hat{\mathbf{B}}\|}
\end{equation}
and the intensity difference $L$
\begin{equation}
L=10 \log_{10}\left(\frac{\left\|\hat{\mathbf{B}}\right\|}{\|\mathbf{B}\|}\right)
\end{equation}
where $\mathbf{B} = [B_0^0, B_1^{-1}, B_1^0, B_1^1, \dots, B_N^{-N}, \dots, B_N^N]^T$
is the coefficient vector,
$\widehat{\mathbf{B}}$ is the estimated HOA coefficients and $\mathbf{B}$ is the ideal HOA coefficients. 

\subsubsection{Signal-to-Distortion Ratio}

The Signal-to-Distortion Ratio (SDR) evaluates the quality of HOA coefficient estimation\citep{2021Estimation}. For each spherical harmonic coefficient of order $n$ and degree $m$, SDR is defined as
\begin{equation}
\operatorname{SDR}\left(B_{n}^m\right)=10 \log_{10} \frac{\left|B_{n}^m\right|^2}{\left|B_{n}^m-\hat{B}_{n}^m\right|^2}
\end{equation}
where $B_{n}^m$ represents the ideal HOA and $\hat{B}_{n}^m$ is its estimated value. This metric provides a coefficient-wise assessment of the estimation accuracy, complementing the global metrics of spatial correlation and intensity difference.

\subsubsection{Pressure Reconstruction Error}
Inspired by \cite{2021Estimation}, the normalized mean square error (NMSE) between the truncated theoretical sound pressure $p_{\text{trunc}}$ and the reconstructed approximated pressure $\hat{p}$ on a spherical surface of radius $R$ is defined as
\begin{equation}
\varepsilon_{\text{error}}(k, R) \coloneqq \frac{\lVert \hat{p} - p_{\text{trunc}} \rVert_S^2}{\lVert p_{\text{trunc}} \rVert_S^2},
\end{equation}
where $\lVert \cdot \rVert_S^2$ represents the squared $L^2$-norm over the spherical surface $S$ with $dS = R^2\sin\theta d\theta d\phi$.

Leveraging spherical harmonic orthogonality, NMSE can be expressed as
\begin{equation}
{
\varepsilon_{\text{error}}(k, R) = \frac{\sum\limits_{n=0}^4 \sum\limits_{m=-n}^n |\hat{B}_n^m - B_n^m|^2 |j_n(kR)|^2}{\sum\limits_{n=0}^4 \sum\limits_{m=-n}^n |B_n^m|^2 |j_n(kR)|^2}.
}
\end{equation}

\begin{figure}[t]
\centering
\includegraphics[width=0.96\linewidth]{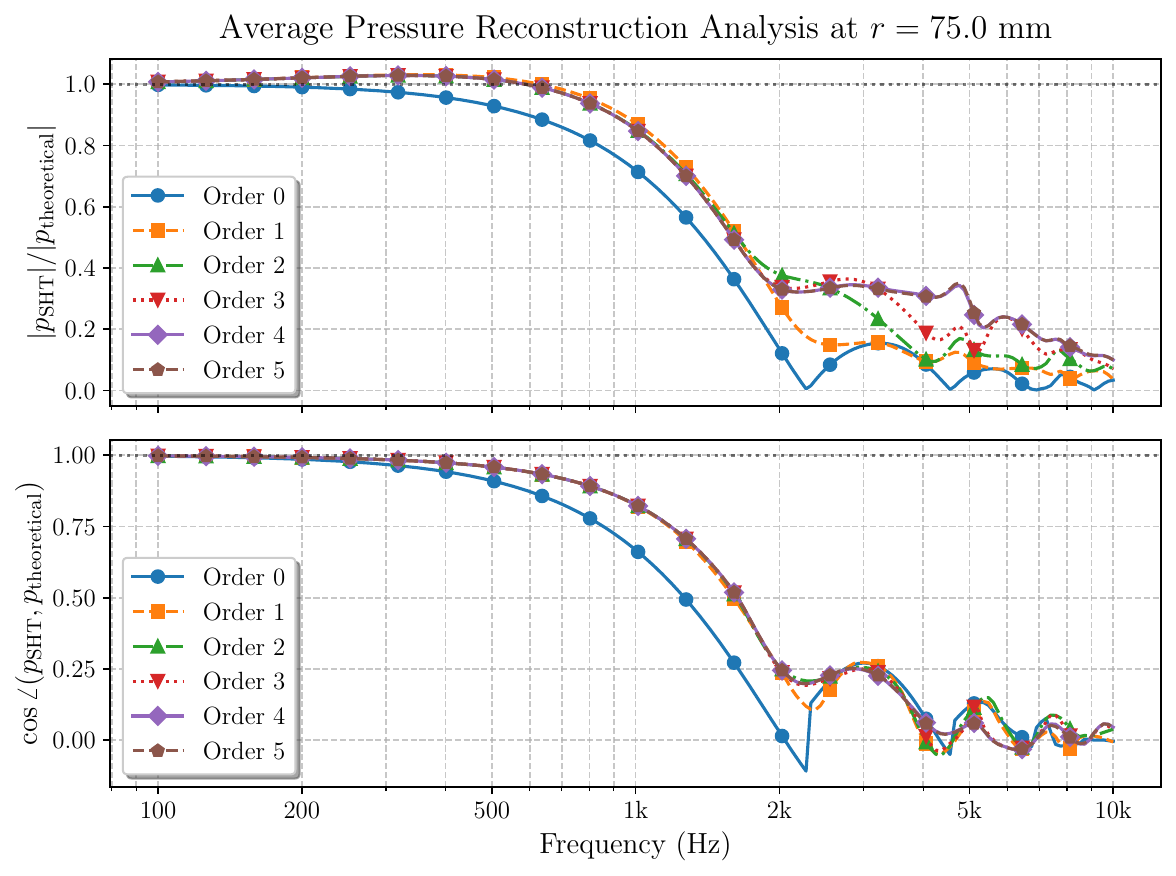}
\caption{Sound field reconstruction accuracy for different upscaling orders. Top: Pressure amplitude ratio \(|p_\mathrm{new}|/|p_\mathrm{theor.}|\); Bottom: Pressure vector cosine similarity \(\cos\angle(p_\mathrm{new}, p_\mathrm{theor.})\).}
\label{fg:upscaling}

\end{figure}
\subsection{Ablation Experiment}
\subsubsection{Order Upscaling}

\begin{figure}[t]
\centering
\includegraphics[width=0.90\linewidth]{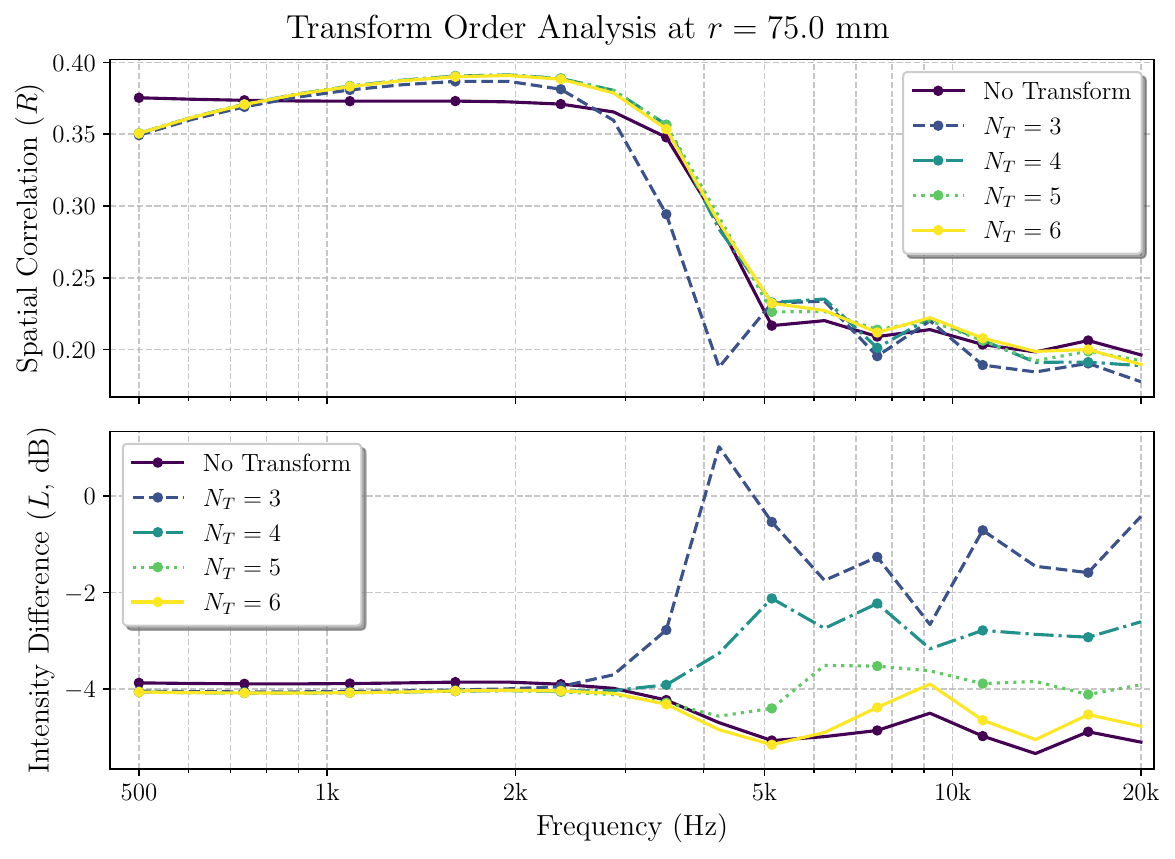}
\caption{Spatial correlation $R$ and intensity difference $L$ between HOA vector $\hat{\mathbf{B}}$ and $\mathbf{B}$ for different transform orders.}
\label{fg:transform_RL}

\end{figure}

As described in Section \ref{subsec:proposed}, beamforming filter $\mathbf{h}_{n_s}^{m_s}(\omega)$ is designed for each spherical harmonic function separately. This architecture ensures mutual independence between filters across upscaling orders—even with limited number of microphones or sparse source directions in the array manifold matrix. This leads us to investigate how our method affects sound field reconstruction as the maximum expansion order increases.

We employ the methodology outlined in Section \ref{subsec:proposed} to an SPMA as depicted in Fig. \ref{fg:arraypos}. Different maximum orders of upscaling are set and the sound pressure is reconstructed on a $r = 0.075\text{m}$ sphere to judge the performance. Positions of reconstructed sound pressure $\theta_{Rec},\phi_{Rec}$ and the directions of arriving plane wave $\theta_{k},\phi_{k}$ are both obtained through Fibonacci sphere uniform sampling\citep{2010Measurement} of one thousand points. We average the corresponding metrics to produce the figures below. Evaluation metrics include the ratio of the reconstructed sound pressure amplitude to the ideal sound pressure amplitude $\langle|p_\mathrm{new}|/|p_\mathrm{theoretical}|\rangle$ and the cosine angle between the reconstructed sound pressure vector and the ideal sound pressure vector $\langle\cos \angle(p_\mathrm{new}, p_\mathrm{theoretical})\rangle$. 

Fig. \ref{fg:upscaling} shows notable improvement through upscaling to order 4. Further upscaling yields negligible improvements. Therefore, we will focus on order 4 as an example for the analysis of HOA coefficient estimation.

\subsubsection{DSHT Order of Array Manifold Matrix}
Next, we need to determine the order of DSHT utilized in Section \ref{subsec:proposed}, incorporating the array manifold matrix into the SHD. Though mathematically the order of DSHT is only limited by the spatial sampling scheme \cite{rafaely2015fundamentals}, which in this section corresponds to the number of steering vectors simulated or measured. In real applications, appropriate truncation helps prevent spatial aliasing in steering vectors, thus improving the following Ambisonics encoding process. As depicted in Fig. \ref{fg:transform_RL}, we compare the spatial correlation and intensity difference for different DSHT orders $N_T=3,4,5,6$ and determine that an appropriate DSHT order is $N=4$, as both $R$ and $L$ metrics show no notable improvement with further increasing $N$. 

\subsection{Simple Condition}
\subsubsection{Simulation Experiments}

In this section, we conduct a detailed comparison between the baseline method in Section \ref{section:Ambisonics} and our method across different metrics. We simulate the theoretical sound pressure at each microphone for a total of $K$ plane wave sound sources with different directions of arrival $\theta_k,\phi_k$, resulting in the steering vector given by
\begin{equation}
\mathbf{d}(\omega, \theta_{k}, \phi_{k})=
[p_{1}(\theta_{k},\phi_{k}), \dots, p_{Q}(\theta_{k}, \phi_{k})]^T.
\end{equation}
Through substituting $\mathbf{d}(\omega, \theta_{k}, \phi_{k})$ into $\mathbf{D}$ and solving Equation \ref{eq:SHDbfEncode} for each spherical harmonic function, the beamforming weight $\mathbf{h}_n^m(\omega)$ is obtained and Equation \ref{HOA calc} can be used for Ambisonics encoding.
\begin{figure}[t]
\centering
\includegraphics[width=0.90\linewidth]{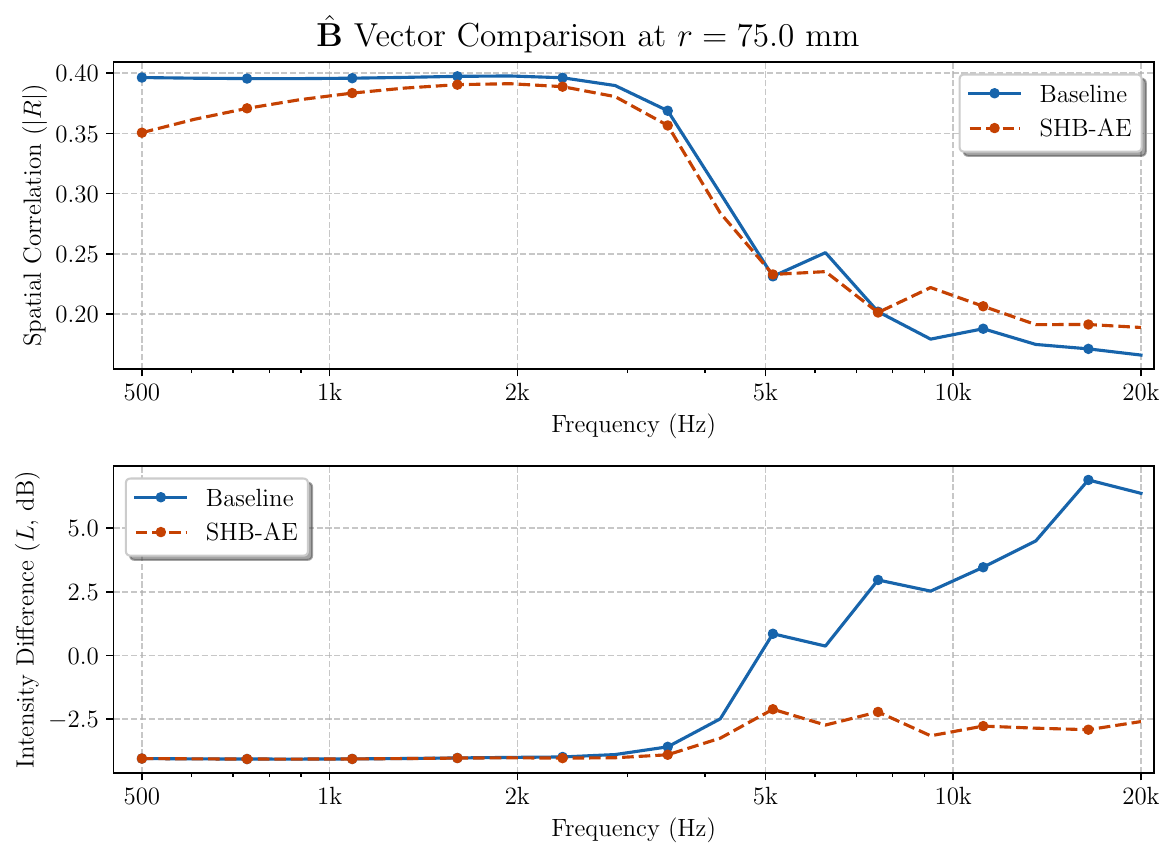}
\caption{$\hat{\mathbf{B}}$ vector estimation comparison between baseline method and our method.}
\label{fg:RL_Simulation}

\end{figure}

\begin{figure}[t]
\centering
\includegraphics[width=0.90\linewidth]{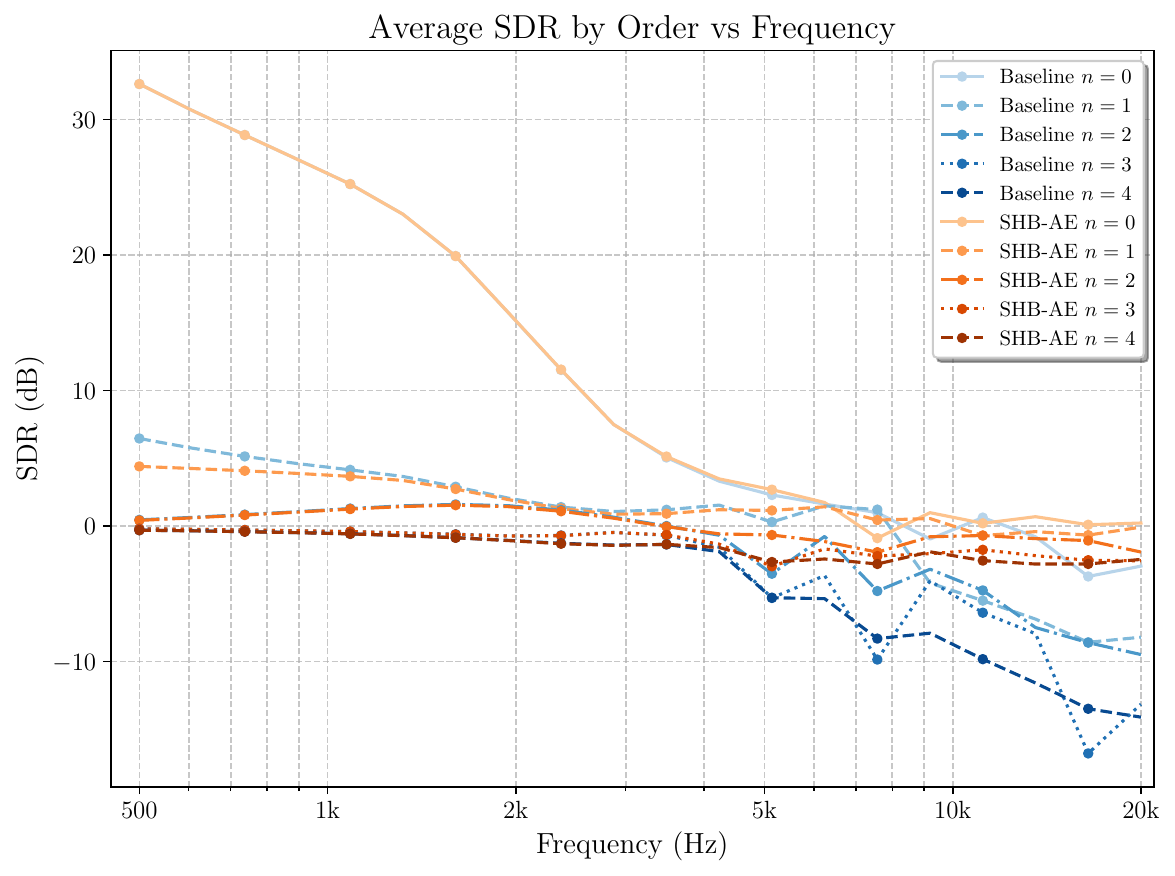}
\caption{Average SDR by order($n$). }
\label{fg:SDR_Simulation}

\end{figure}
As illustrated in Fig. \ref{fg:RL_Simulation}, both the baseline and our method exhibit similar estimates of the vector angles in the overall HOA estimation. However, the baseline begins to exhibit significant amplitude deviations in the mid-to-high frequency range. It is also reflected in Fig. \ref{fg:SDR_Simulation}, where the SDR metric of the baseline method is not only significantly lower than that of our approach in terms of overall mean but also displays a more pronounced downward trend in the mid to high frequencies, indicating its divergence from the theoretical HOA coefficients. We believe that the reason for the phenomenon lies on the aliasing frequency of the SPMA, indicating the difficulty of encoding high frequency HOA. By calculating the half-wavelength frequencies corresponding to the distances between microphones in our setup, we find that four out of the six distances correspond to frequencies below 2 kHz, resulting in significant spatial aliasing in the high-frequency range. In contrast, our method focuses on the number of sound source directions (steering vectors) rather than the number of microphones, making our system of linear equations overdetermined. Furthermore, as the equations are transformed into the spherical harmonic domain (SHD), our method exhibits improved robustness.

\begin{figure}[t]
\centering
\includegraphics[width=0.96\linewidth]{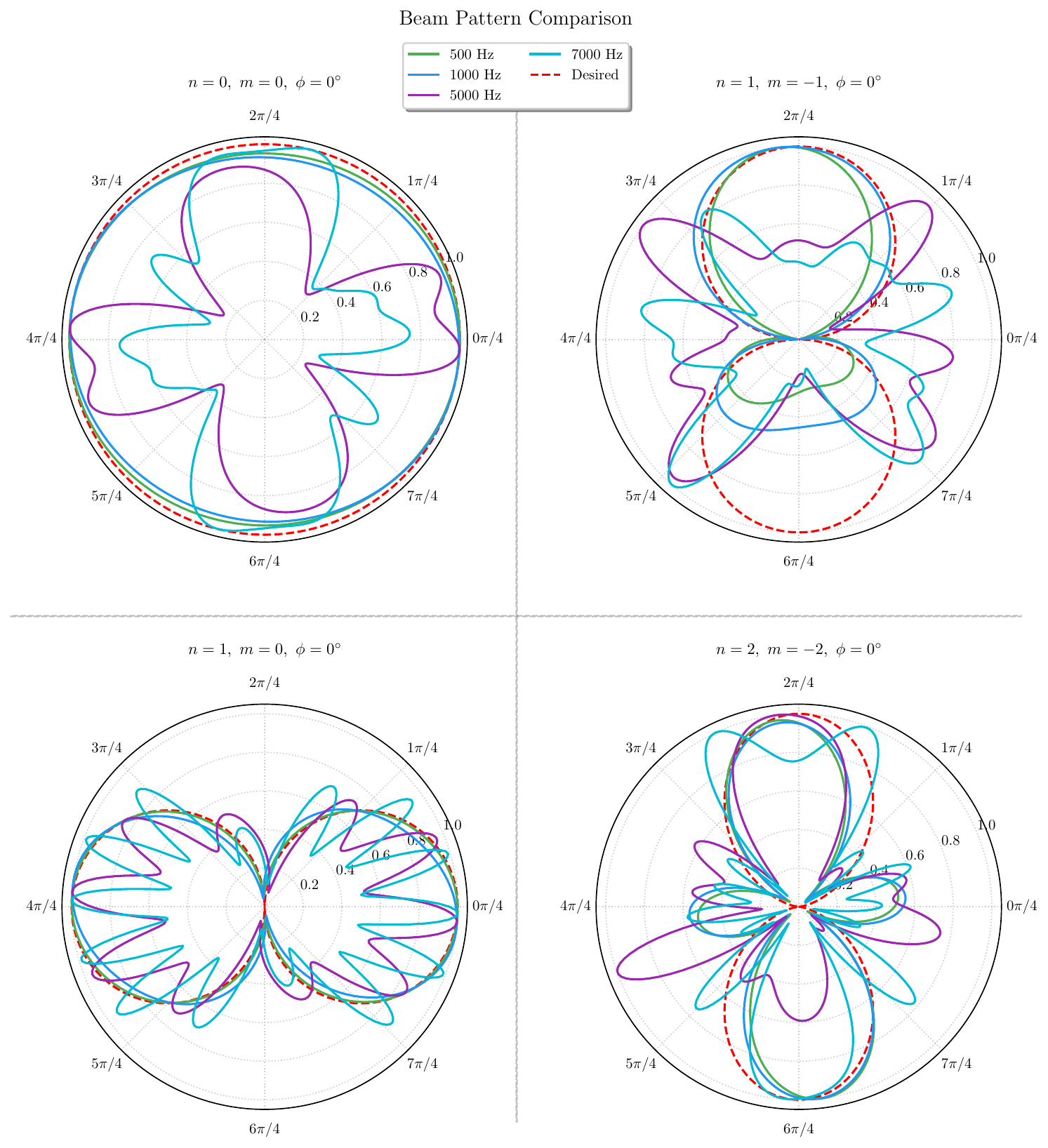}
\caption{Beam pattern examples. $n, m$ denotes the order and degree of the harmonic function, while $\phi$ denotes the azimuth angle.}
\label{fg:beam_pattern}

\end{figure}

We can also judge the estimation of the HOA coefficients and beamforming in a more intuitive way. Fig. \ref{fg:beam_pattern} presents beam patterns for different spherical harmonic orders at various frequencies within the same plane, illustrating that our method effectively achieves the task of mitigating the spherical harmonics through beamforming. Furthermore, the beam patterns at low frequencies are generally smoother, while rougher at higher frequencies, because of the constraints corresponding to the fixed source direction becoming more pronounced, indicating a weakening of the smoothing effect of spherical harmonic transformations on beam formation. This phenomenon can also be attributed to high-frequency aliasing caused by insufficient element spacing. Additionally, it is a plausible theoretical explanation that the low-order spherical harmonic basis functions exhibit slow spatial variation, rendering them incapable of adequately expressing the details of high-frequency acoustic fields. 

It is important to note that both the baseline method and our approach have limited frequency-dependent characteristics in their least squares solutions, which cannot be simply attributed to the high condition number at high frequencies. In fact, the condition numbers of the two linear systems fluctuate with frequency and do not exhibit a clear monotonicity.

\subsubsection{Practical Experiments}

\begin{figure}[t]
\centering
\includegraphics[width=0.92\linewidth]{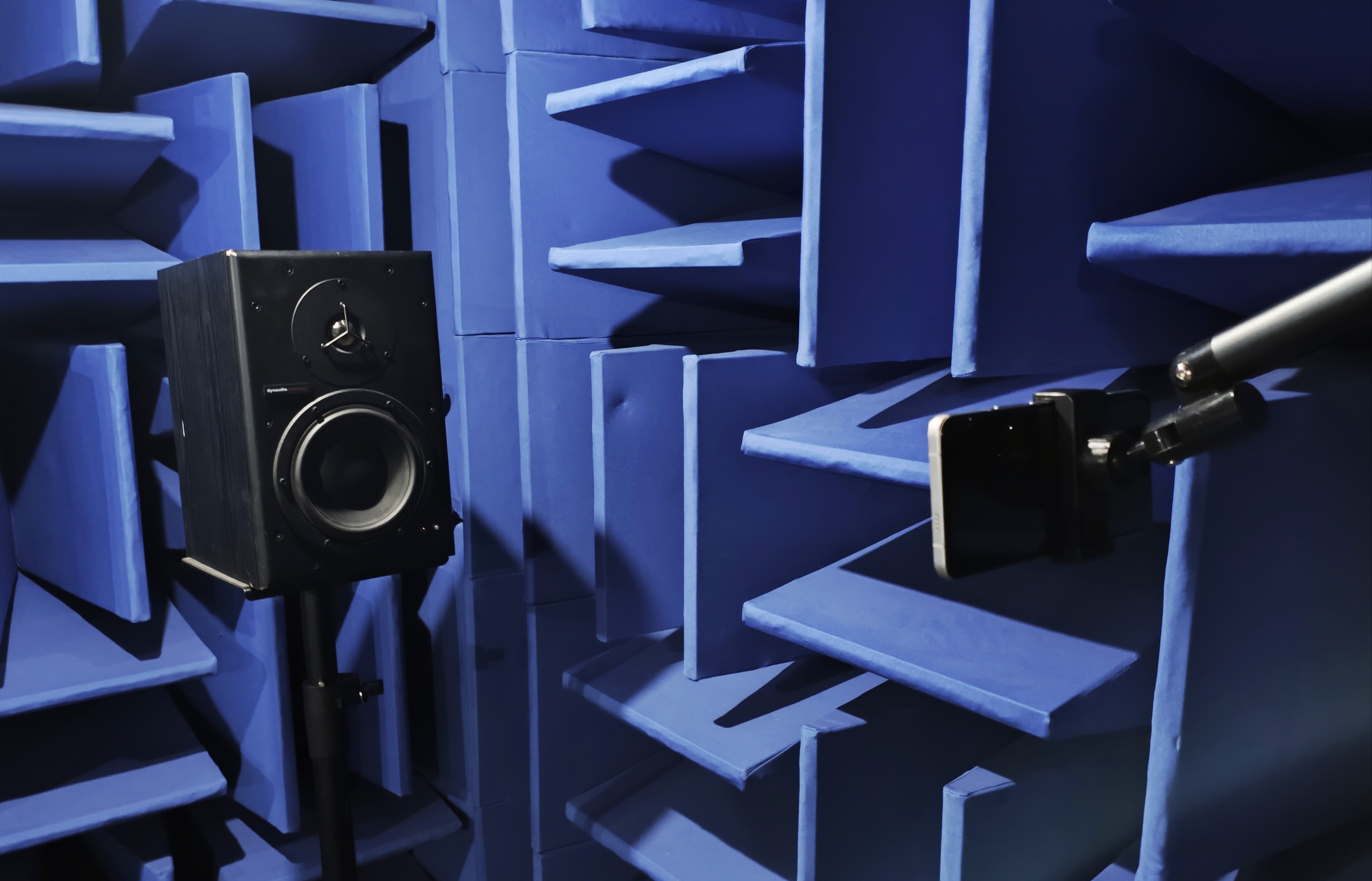}
\caption{The acoustic system utilized for measuring the impulse responses of the SPMA}
\label{fg:chamber}

\end{figure}

\begin{figure}[t]
\centering
\includegraphics[width=0.96\linewidth]{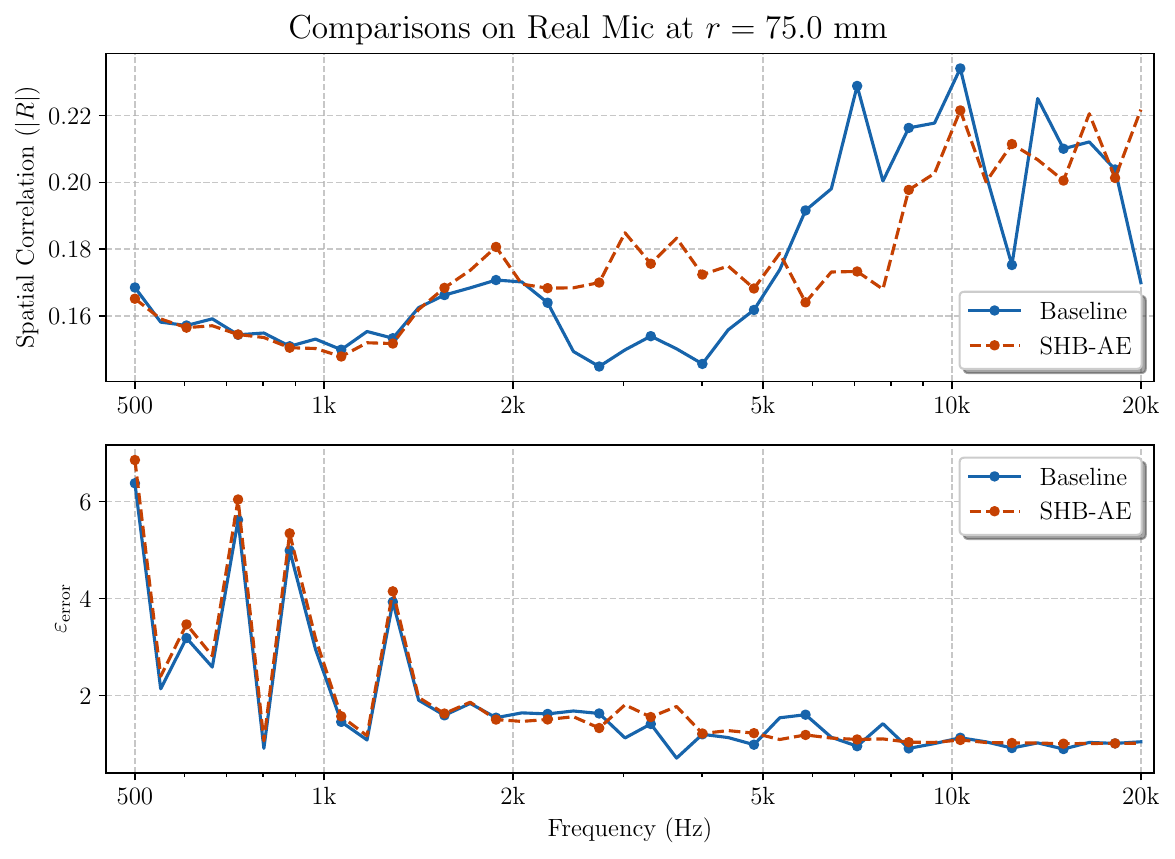}
\caption{$R$ and $\varepsilon_{\mathrm{error}}$ evaluation on real microphone arrays.}
\label{fg:Realmic}

\end{figure}
\begin{figure}[t]
\centering
\includegraphics[width=0.90\linewidth]{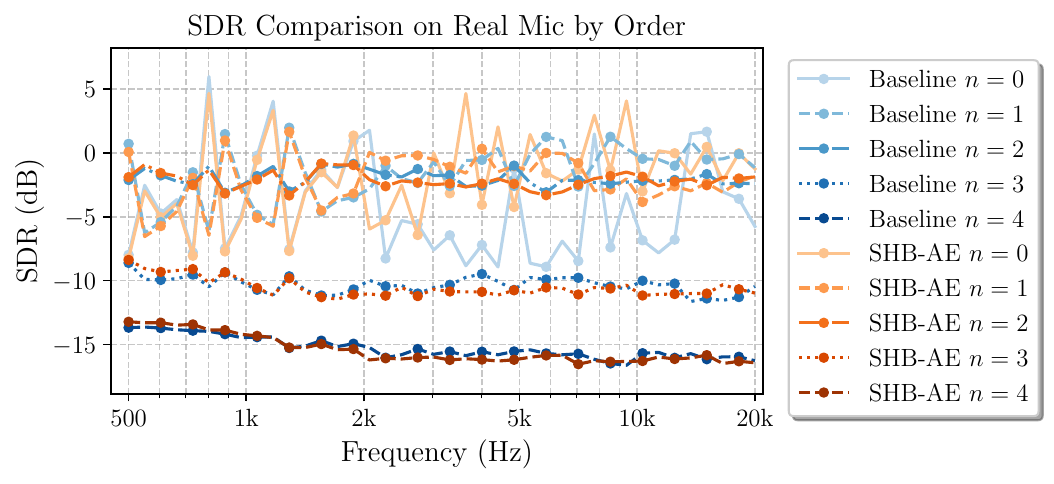}
\caption{SDR evaluation on real microphone arrays.}
\label{fg:Realmic_SDR}

\end{figure}

We also conduct experiments using data collected from real microphones. In the anechoic chamber depicted in Fig. \ref{fg:chamber}, impulse responses of the SPMA are measured through exponential sine sweep signals exciting a loudspeaker \cite{farina2007advancements}.

We obtain the sound pressure vector through convolution of the impulse responses with sound source signals. In simulation, the baseline construction of the encoding matrix relies solely on theoretical calculations from Equation \ref{eq:A_Encode_thr}, which fails to utilize the manifold information from the real SPMA, thus preventing effective adjustment of actual sound pressure encoding.

Therefore, in our actual experiments, the baseline method is considered to further separate the radial and angular components in the encoding matrix $A$, which is essentially equivalent to the beamforming assumption we introduced. Our method performs DSHT and frequency division operations based on this assumption, and our actual experiments focus on this comparison.
Figures \ref{fg:Realmic} and \ref{fg:Realmic_SDR} demonstrate the performance metrics in our physical experiments. We set $\omega_{th}=2\mathrm{kHz}$ by experiments. Due to the influence of various external factors on the actual collected data, we performed a uniform energy normalization when calculating the HOA coefficients. This normalization renders the \( L \) metric ineffective, but does not affect the evaluation results of other metrics.

The experimental results shows that our method achieves significant improvements in spatial correlation within the 2-5 kHz range, accompanied by a notable reduction in reconstruction error. In the high-frequency band, the frequency-divided curves exhibit relatively stable behavior, while the baseline method shows substantial fluctuations. These findings confirm that our frequency division strategy above 2 kHz achieves its intended purpose—maintaining frequency consistency is more effective when withholding phase information rather than providing incorrect phase information. SHT also contributes to this high-frequency stability characteristic, which is well reflected in the SDR metrics.
\subsection{Complex Conditions}

Furthermore, we conduct additional simulation and practical experiments in different noise levels and reverberation time scenarios to judge the robustness of methods. Consistent with the previous sections, in the simulation experiments, we compared the results of the baseline with those obtained through beamforming and spherical harmonic transformation. In the practical experiments, we compared the outcomes of beamforming alone with those of spherical harmonic transformation coupled with frequency division.
\subsubsection{Noise}
\begin{table}[t]
\centering
\small
\caption{Simulation results under noise with different SNR(dB).}
\begin{adjustbox}{max width=\columnwidth}
\begin{tabular}{cc|cccc}
\hline
$\text{SNR}$ & Method & $\varepsilon_{\mathrm{error}}\downarrow$ & $R\uparrow$ & $L\rightarrow0$ & $\text{SDR}\uparrow$ \\
\hline
\multirow{2}{*}{30} & Baseline & 1.90 & \textbf{0.32} & \textbf{-1.38} & -1.74 \\
& SHB-AE & \textbf{0.70} & {0.30} & {-3.43} & \textbf{0.02} \\
\hline
\multirow{2}{*}{20} & Baseline & 1.98 & \textbf{0.29} & \textbf{-0.76} & -2.29 \\
& SHB-AE & \textbf{0.75} & 0.28 & {-2.82} & \textbf{-0.46} \\
\hline
\multirow{2}{*}{10} & Baseline & 2.79 & 0.25 & \textbf{0.71} & -3.45 \\
& SHB-AE & \textbf{1.29} &\textbf{0.25} & -1.36 & \textbf{-1.52} \\
\hline
\multirow{2}{*}{0} & Baseline & 10.90 & 0.20 & 3.50 & -5.97 \\
& SHB-AE & \textbf{6.69} & \textbf{0.20} & \textbf{1.42} & \textbf{-3.59} \\
\hline
\end{tabular}
\end{adjustbox}
\end{table}

\begin{table}[t]
\centering
\small
\caption{Practical experimental results under noise with different SNR(dB).}
\begin{adjustbox}{max width=\columnwidth}
\begin{tabular}{cc|ccc}
\hline
SNR & Method & $\varepsilon_{\mathrm{error}}\downarrow$ & $R\uparrow$ & $\text{SDR}\uparrow$ \\
\hline
\multirow{2}{*}{30.0} & Baseline & \textbf{2.02} & 0.18 & \textbf{-9.53} \\
& SHB-AE & 2.16 & \textbf{0.18} & -9.54 \\
\hline
\multirow{2}{*}{20.0} & Baseline & \textbf{1.86} & 0.18 & \textbf{-9.49} \\
& SHB-AE & 1.99 & \textbf{0.18} & -9.52 \\
\hline
\multirow{2}{*}{10.0} & Baseline & \textbf{1.66} & 0.18 & \textbf{-9.39} \\
& SHB-AE & 1.76 & \textbf{0.19} & -9.42 \\
\hline
\multirow{2}{*}{0.0} & Baseline & \textbf{1.63} & \textbf{0.19} & -9.37 \\
& SHB-AE & 1.71 & 0.18 & \textbf{-9.36} \\
\hline
\end{tabular}
\end{adjustbox}
\end{table}
To test the robustness of methods against additive, spatially white Gaussian noise under different Signal-to-Noise Ratio (SNR) is utilized. The experimental results demonstrate that the proposed method exhibits strong robustness across various noise levels. In simulation experiments, as SNR decreases from 30 dB to 0 dB, the proposed method consistently outperforms the baseline method in terms of $\varepsilon_{\text{error}}$, particularly under low SNR conditions (0 dB), where the proposed method's error (6.69) is significantly lower than that of the baseline method (10.90), indicating its excellent noise resistance capability. Regarding the correlation coefficient $R$, both methods show similar performance that $R$ declines as the SNR decreases as expected. In terms of the SDR, the proposed method outperforms the baseline across all SNR levels, indicating superior signal fidelity.

In practical experiments, the performance difference between the proposed method and traditional beamforming techniques is not as obvious, suggesting that the introduction of SHT and frequency division only leads to limited advantages. In fact, results of different frequencies fluctuate significantly for both methods but the reason remains unknown. The comparable measurement metrics indicate that both methods exhibit similar noise resistance performance in practical applications.
\subsubsection{Reverberation}

\begin{table}[t]
\centering
\small
\caption{Simulation results under reverberation with different RT60(s).}
\begin{adjustbox}{max width=\columnwidth}
\begin{tabular}{cc|cccc}
\hline
$\text{RT60}$ & Method & $\varepsilon_{\mathrm{error}}\downarrow$ & $R\uparrow$ & $L\rightarrow0$ & $\text{SDR}\uparrow$ \\
\hline
\multirow{2}{*}{0.2} & Baseline & 3.36 & 0.21 & \textbf{-10.29} & -0.50 \\
& SHB-AE & \textbf{1.89} & \textbf{0.21} & -12.30 & \textbf{-0.29} \\
\hline
\multirow{2}{*}{0.5} & Baseline & 156.53 & 0.21 & \textbf{-1.65} & -2.78 \\
& SHB-AE & \textbf{48.39} & \textbf{0.22} & -3.75 & \textbf{-1.27} \\
\hline
\multirow{2}{*}{1.0} & Baseline & 551.44 & 0.22 & \textbf{0.91} & -4.61 \\
& SHB-AE & \textbf{163.68} & \textbf{0.22} & -1.20 & \textbf{-2.26} \\
\hline
\multirow{2}{*}{2.0} & Baseline & 1020.52 & 0.22 & 2.16 & -5.75 \\
& SHB-AE & \textbf{300.26} & \textbf{0.22} & \textbf{0.04} & \textbf{-2.99} \\
\hline
\end{tabular}
\end{adjustbox}
\end{table}

\begin{table}[t]
\centering
\small
\caption{Practical experimental results under reverberation with different RT60(s).}
\begin{adjustbox}{max width=\columnwidth}
\begin{tabular}{cc|ccc}
\hline
RT60 & Method & $\varepsilon_{\mathrm{error}}\downarrow$ & $R\uparrow$ & $\text{SDR}\uparrow$ \\
\hline
\multirow{2}{*}{0.2} & Baseline & 2.39 & \textbf{0.20} & -9.91 \\
& SHB-AE & \textbf{1.75} & 0.19 & \textbf{-9.40} \\
\hline
\multirow{2}{*}{0.5} & Baseline & 2.00 & \textbf{0.19} & -9.83 \\
& SHB-AE & \textbf{1.79} & 0.18 & \textbf{-9.45} \\
\hline
\multirow{2}{*}{1.0} & Baseline & 1.89 & \textbf{0.19} & -9.81 \\
& SHB-AE & \textbf{1.83} & 0.18 & \textbf{-9.46} \\
\hline
\multirow{2}{*}{2.0} & Baseline & 1.86 & \textbf{0.19} & -9.81 \\
& SHB-AE & \textbf{1.85} & 0.18 & \textbf{-9.46} \\
\hline
\end{tabular}
\end{adjustbox}
\end{table}

In the experiments conducted under reverberation conditions, a typical 3D shoebox room with different Reverberation Time 60 (RT60) is utilized. After simulating the room impulse responses with pyroomacoustics \cite{scheibler2018pyroomacoustics}, convolution between them and simulated or measured steering vectors is done for mitigating the room reverberation nature. 

The simulation results demonstrate that system performance degrades with increasing RT60, primarily manifested in the significant rise of the error metric $\varepsilon_{\mathrm{error}}$. However, compared to the baseline method, the proposed approach exhibits superior robustness against reverberation. Under mild reverberation conditions (RT60=0.2s), the proposed method reduces the error from 3.36 to 1.89 and in highly reverberant environments (RT60=2.0s), the proposed method maintains the error at 300.26 compared to 1020.52 of the baseline method, showcasing substantial performance advantages.

In practical experiments, although the overall error levels are lower than simulation results due to normalization, our method consistently maintains its performance advantages. Specifically, across all RT60 conditions, our method outperforms the beamforming approach in both $\varepsilon_{\mathrm{error}}$ and SDR metrics. Notably, the impact of RT60 variations on system performance in real environments is considerably less pronounced than observed in simulations, which may be attributed to the more complex acoustic characteristics in practical settings. Additionally, the $R$-value remains relatively stable across different conditions, indicating its insensitivity to reverberant conditions.

\section{Summary}
For the task of Ambisonics encoding and upsacling on SPMA, we propose SHB-AE, a spherical harmonic beamforming-based method which designs beamformers for each spherical harmonic function based on the array manifold and introducing DSHT together with frequency division operations. Experimental validation was conducted on both simulated and real SPMA across varying reverberant and noisy environments. Results demonstrated that our method achieved superior performance in various metrics and successfully encodes and up-scales HOA. Its robustness in high-frequency ranges and complex conditions is also obvious. This approach offers a practical solution for spatial audio recording using portable devices, potentially enabling wider adoption of immersive audio capture in AR/VR applications.

\section*{Acknowledgment}
This work is supported by the National Key Research and Development Program of China (No.2024YFB2808902), and the High-performance Computing Platform of Peking University.

\bibliographystyle{IEEEbib}
\bibliography{refs}

\end{document}